\documentclass[twocolumn,secnumarabic,amssymb,nobibnotes,aps,prl]{revtex4-1}
\usepackage{color}
\usepackage{graphicx}
\begin{document}

\title{Lattice strain accompanying the colossal magnetoresistance effect in EuB$_6$}

\author{Rudra Sekhar Manna}
\author{Pintu Das}
\altaffiliation[Present address: ]{Dept.\ of Physics, Indian Institute of Technology, New Dehli, India}
\author{Mariano de Souza}
\altaffiliation[Present address: ]{IGCE, Unesp.-Univ.\ Estadual Paulista, Dept.\ de F\'{i}sica, Rio Claro, Brazil}
\author{Michael Lang}
\author{Jens M\"uller}
\email[Email: ]{j.mueller@physik.uni-frankfurt.de}
\affiliation{Institute of Physics, Goethe-University Frankfurt, 60438 Frankfurt (M), SFB/TR49, Germany}
\author{Stephan von Moln\'{a}r}
\affiliation{Department of Physics, Florida State University, Tallahassee, Florida 32306, USA}
\author{Zachary Fisk}
\affiliation{Department of Physics, University of California, Irvine, California 92697, USA}


\date{\today}

\begin{abstract}
The coupling of magnetic and electronic degrees of freedom to the crystal lattice in the ferromagnetic semimetal EuB$_6$, which exhibits a complex ferromagnetic order and a colossal magnetoresistance (CMR) effect, 
is studied by high-resolution thermal expansion and magnetostriction experiments. EuB$_6$ may be viewed as a model system, where pure magnetism-tuned transport and the response of the crystal lattice can be studied in a comparatively simple environment, {\it i.e.}, not influenced by strong crystal-electric field effects and Jahn-Teller distortions. We find a very large lattice response, quantified by (i) the magnetic Gr\"uneisen parameter, (ii) the spontaneous strain when entering the ferromagnetic region and (iii) the magnetostriction in the paramagnetic temperature regime. Our analysis reveals that a significant part of the lattice effects originates in the magnetically-driven delocalization of charge carriers, consistent with the scenario of percolating magnetic polarons. A strong effect of the formation and dynamics of local magnetic clusters on the lattice parameters is suggested to be a general feature of CMR materials.
\end{abstract}

\maketitle

Materials, in which the resistivity exhibits drastic changes in response to an external magnetic field, are of great interest both from a fundamental as well as a technological point of view. Those anomalous magneto-transport effects are particularly strongly pronounced close to a combined magnetic and insulator-metal transition, where a large or even a colossal magnetoresistance (CMR) can be observed. Prominent examples include magnetic semiconductors, rare-earth chalcogenides, silicides and hexaborides, Mn-based pyrochlors, as well as the mixed-valent rare-earth manganites \cite{Molnar1967,RamirezScience1997,CNRRao1998,KaminskiPRL2002,Li2012}.
One route for describing the CMR effect involves the formation of magnetic polarons (MPs).
This concept was first suggested based on experiments on Eu$_{1-x}$Gd$_x$Se \cite{Molnar1967} and shortly thereafter given a theoretical foundation \cite{Kasuya1968}. 
MPs are formed when it is energetically favorable for the charge carriers to localize and spin polarize the surrounding local moments over a finite distance. 
With increasing magnetic field, these ordered clusters may grow in size, accompanied by a progressive alignment of
the spins outside the ordered clusters thereby facilitating the charge transport.  
The existence of magnetic clusters (tantamount to MPs) in some manganites has been demonstrated by a concomitant lattice distortion, the field dependence of which closely follows the magnetoresistivity \cite{TeresaNature1997}.
Despite considerable efforts to understand the interplay between spin, charge and lattice degrees of freedom in the CMR effect for the various materials, see e.g.\ \cite{IbarraPRL1995,TeresaNature1997,GarciaLandaPRL2000,Downward2005,Souza2005,Zhou2008}, no general 
picture has evolved yet. For the manganites, in particular, the reason for that may be related to their complexity due to the simultaneous action of strong crystal-electric field (CEF) and Jahn-Teller (JT) effects.


Here, we present a detailed study of lattice effects accompanying the CMR effect in the comparatively simple system EuB$_{6}$, which -- in first order -- is devoid of CEF and JT effects. This material has a body-centered cubic structure where B$_{6}$ octahedra are surrounded by eight Eu metal atoms residing at the corners of a cube. Due to the Eu$^{2+}$ Hund's rule ground state configuration of $^{8}$S$_{7/2}$, which is magnetically isotropic, EuB$_{6}$ may be viewed as a model system for studying purely spin-tuned transport phenomena. Despite its simplicity, the system shows a rich phenomenology. The inverse magnetic susceptibility, $\chi^{-1}$, of the Eu moments shows a linear temperature dependence for $T \gtrsim 20$\,K with a paramagnetic (PM) Curie temperature of $\Theta_p \approx 15.6$\,K \cite{FiskJAP1979,ZhangPRL2009} and the FM state is reached via two consecutive transitions at $T_{c_1} = 15.3$\,K and $T_{c_2} = 12.6$\,K \cite{SuellowPRB1998,SuellowPRB2000,Urbano2004,Das2012}. Moreover, the PM-FM transition is accompanied by a drastic reduction of the resistance at zero applied field as well as a CMR effect.
It has been proposed that the large negative MR in EuB$_{6}$ at $T_{c_1}$ is a percolation-type transition resulting from the overlap of MPs, which causes a delocalization of the hole carriers \cite{NyhusPRB1997,SuellowPRB2000,ZhangPRL2009}. 
Upon cooling, the polaronic clusters percolate at $T \leq T_{c_1}$, and finally merge at $T \leq T_{c_2}$, where bulk FM order sets in -- a scenario in accordance with recent magnetic \cite{BrooksPRB2004} and transport data \cite{ZhangPRL2009,Das2012}.\\
Given the pronounced lattice distortions accompanying the formation of MPs in the manganites, 
it is natural to look for similar effects also in EuB$_{6}$. To the best of our knowledge, only limited information on the lattice effects in EuB$_{6}$ has been available so far \cite{SuellowPRB1998,Sirota2000,ZherlitsynEPJB2001,MartinhoJPCM2009}. 

In this work, we report on high-resolution thermal expansion and magnetostriction measurements of EuB$_6$. The salient result of our study is the observation of a strong lattice strain disclosing a clear correspondence with the material's CMR effect.  
%

Single crystals of EuB$_6$ were grown from Al flux as described in Ref.\ \cite{FiskJAP1979}. For the thermal expansion and magnetostriction measurements, a high-resolution capacitive dilatometer (built after \cite{Pott1983}) was used, enabling the detection of 
length changes $\Delta l \geq 10^{-2}\,{\rm \AA}$. The coefficient of thermal expansion $\alpha(T) = {\rm d}\ln{l}/{\rm d}T$ and the magnetostriction coefficient $\lambda = {\rm d}\ln{l}/{\rm d}B$ were measured along a principal direction of the cubic structure and parallel to the applied magnetic field. The experiments were carried out on two single crystals from different batches yielding similar results. Samples \#1 and \#2 have dimensions $5 \times 1 \times 0.75$\,mm$^3$ and $1 \times 0.5 \times 0.2$\,mm$^3$, respectively.

\begin{figure}[h]
\includegraphics[width=\columnwidth]{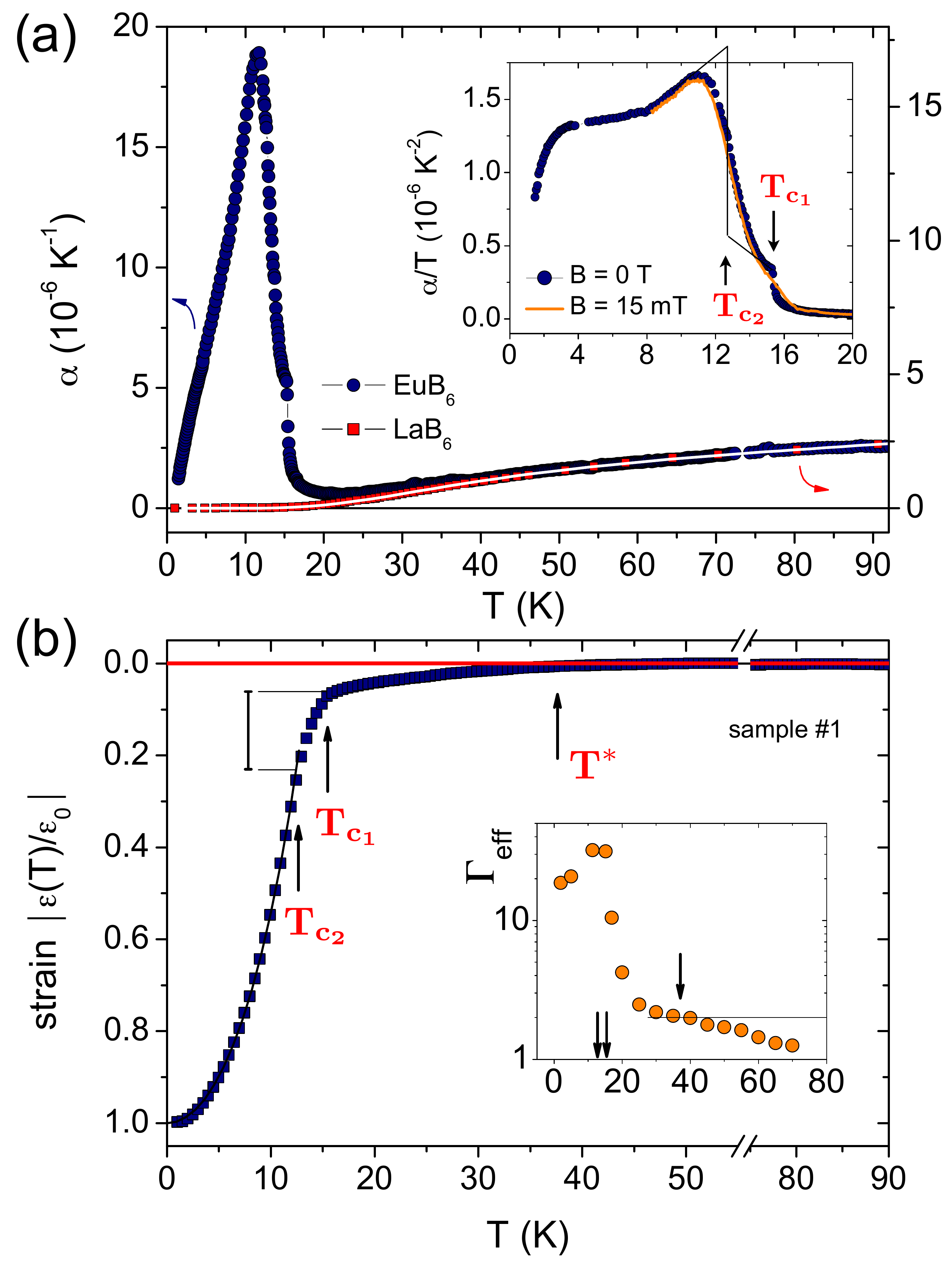}
\caption{\label{Fig1}(Color online) (a) Coefficient of thermal expansion, $\alpha(T)$, {\it vs.}\ $T$, of EuB$_6$ (blue circles, left scale) and LaB$_6$ (red squares, right scale). Inset shows the same data in a plot $\alpha/T$ {\it vs.}\ $T$ together with data at a field of $B = 15$\,{\rm mT}.
White line represents the estimated phonon contribution for EuB$_6$ as described in the text.
(b) Modulus of the (negative) spontaneous strain $\epsilon = (\Delta l/l)_{\rm EuB_6} - (\Delta l/l)_{\rm ph}$ {\it vs.}\ $T$ normalized to the value at zero temperature $\epsilon_0$.
Arrows indicate the transition temperatures $T_{c_1}$ and $T_{c_2}$ (scale bar indicates the strain in this temperature interval) as well as the characteristic temperature for MP formation $T^\ast$. 
Inset shows the effective Gr\"uneisen parameter, $\Gamma_{\rm eff}$, 
as described in the text (arrows mark $T^\ast$, $T_{c_1}$ and $T_{c_2}$).} 
\end{figure}
Figure\,\ref{Fig1}(a) shows the thermal expansion coefficient $\alpha(T)$ of EuB$_6$ and the isostructural LaB$_6$ \cite{LaB6}. Since the $4f$ shell of La$^{3+}$ is empty, LaB$_6$ may serve as a nonmagnetic reference system to EuB$_6$. 
Consequently, the very large positive contribution to the expansivity for $T \lesssim$ 15\,K ($\backsimeq$ $T_{c_1}$), corresponding to a strong contraction of the lattice upon cooling through the FM transition, see Fig.\,\ref{Fig1}(b), is of magnetic origin.
As observed also for the electrical resistivity, we find two subsequent anomalies in the lattice expansivity. In $\alpha(T)/T$ {\it vs.}\ $T$, see inset of Fig.\,\ref{Fig1}(a), there is a sharp transition at $T_{c_1} = 15.4$\,K, corresponding to the $\lambda$-shaped anomaly found in the specific heat \cite{SuellowPRB1998,Urbano2004} and 
a large maximum slightly below 12\,K.
From an entropy-conserving 
equal-areas construction in $\alpha(T)/T$ {\it vs.}\ $T$, also shown in the inset of Fig.\,\ref{Fig1}(a), we find $T_{c_2} = 12.6$\,K \cite{comment0}. We note that two separate anomalies can only be observed in samples of highest quality \cite{SuellowPRB1998}.
In order to determine the magnetic contribution, $\alpha_{\rm mag}$, to the expansivity of EuB$_6$ as accurately as possible, we proceed as follows. Given that at high enough temperatures ($\sim 90$\,K) the expansivities of EuB$_6$ and LaB$_6$ represent, to a good approximation, pure lattice effects, we multiply the expansivity of LaB$_6$, $\alpha_{\rm LaB_6}$, by a factor $k$ such that it matches the expansivity of EuB$_6$ at 90\,K. This factor may account for differences in the lattice anharmonicity of both compounds resulting from the different atomic masses of La and Eu. In fact by using $k = 0.9$ both data sets collapse over a wide range of temperatures $40\,{\rm K} \lesssim T \lesssim 100$\,K (the highest temperature in the present experiment), consistent with $k \alpha_{\rm LaB_6}$($T$) being a good measure of the phonon contribution, $\alpha_{\rm ph}(T)$, of EuB$_6$ \cite{comment01}.\\ 
In Fig.\,\ref{Fig1}(b) we show the relative length change, $\Delta l/l$, of EuB$_6$ as the difference $\epsilon(T) = (\Delta l/l)_{\rm EuB_6} - (\Delta l/l)_{\rm ph}$, corresponding to a spontaneous strain, normalized to the extrapolated value at zero temperature $\epsilon_0$. 
Our data uncover the onset of the negative strain at a temperature around $T^\ast \sim 35 - 40$\,K. This is about the same temperature below which indications for bound MPs have been observed \cite{NyhusPRB1997,SuellowPRB2000,BrooksPRB2004,Das2012}, suggesting that their formation is accompanied by a lattice distortion. 
Upon further cooling, the lattice contraction strongly increases at the percolation transition temperature $T_{c_1}$, and then displays an order paramater-like behavior below $T_{c_2}$ (solid line).
Furthermore, we find that about $15 - 20$\,\% of the spontaneous lattice contraction occurs between $T_{c_1}$ and $T_{c_2}$, see the scale bar in Fig.\,\ref{Fig1}(b),
which, remarkably, is also the amount of Eu moments that already order at $T_{c_1}$ before bulk FM order sets in \cite{SuellowPRB2000}.\\
%
Indications for an anomalous contribution to the expansivity below $T^\ast$ can be found also by looking at the effective Gr\"uneisen parameter $\Gamma_{\rm eff} = (V_{\rm mol}/\kappa_T ) [3\alpha/C_V]$ shown in the inset of Fig.\,\ref{Fig1}(b). Here $V_{\rm mol} = 43.9\,{\rm cm^3/mol}$ is the molar volume, $\kappa_T = 0.6 \times 10^{-11}$\,Pa$^{-1}$ the isothermal compressibility of EuB$_6$ \cite{Lundstrom1982}, $\beta = 3 \alpha$ the volume expansion coefficient and $C_V$ the specific heat \cite{Fujita1980,SuellowPRB1998,Urbano2004}. The so-derived $\Gamma_{\rm eff} = \sum_i \Gamma_i {C_{V}}_i /\sum_i {C_{V}}_i$ consists of contributions from each subsystem $\Gamma_i$, such as $i =$ lattice, electronic and magnetic, weighted by its specific heat ${C_{V}}_i$. $\Gamma_{\rm eff}$ is usually of order unity for simple metals or insulators \cite{Barron1980}.
As shown in the inset of Fig.\,\ref{Fig1}(b), $\Gamma_{\rm eff}$ is about 1 for $T > 70$\,K and gradually increases upon cooling, reaching an enhanced value of $\Gamma_{\rm eff} \sim 2$ around 40\,K (close to $T^\ast$), where it shows a small plateau. Hence, far above $T^\ast$, in the PM regime, $\Gamma_{\rm eff}$ is consistent with a phonon-dominated anharmonicity. In the regime of isolated MPs, $T_{c_1} \leq T \leq T^\ast$, we observe an increase from 2 at 40\,K to $\Gamma_{\rm eff} \sim 5$ at around $T = 20$\,K. Finally, the drastic enhancement of $\Gamma_{\rm eff}$ on approaching the FM regime for $T < 20$\,K is strongly influenced by percolating MPs and charge delocalization, as we will argue below.\\

\begin{figure}[htb]
\includegraphics[width=0.8\columnwidth]{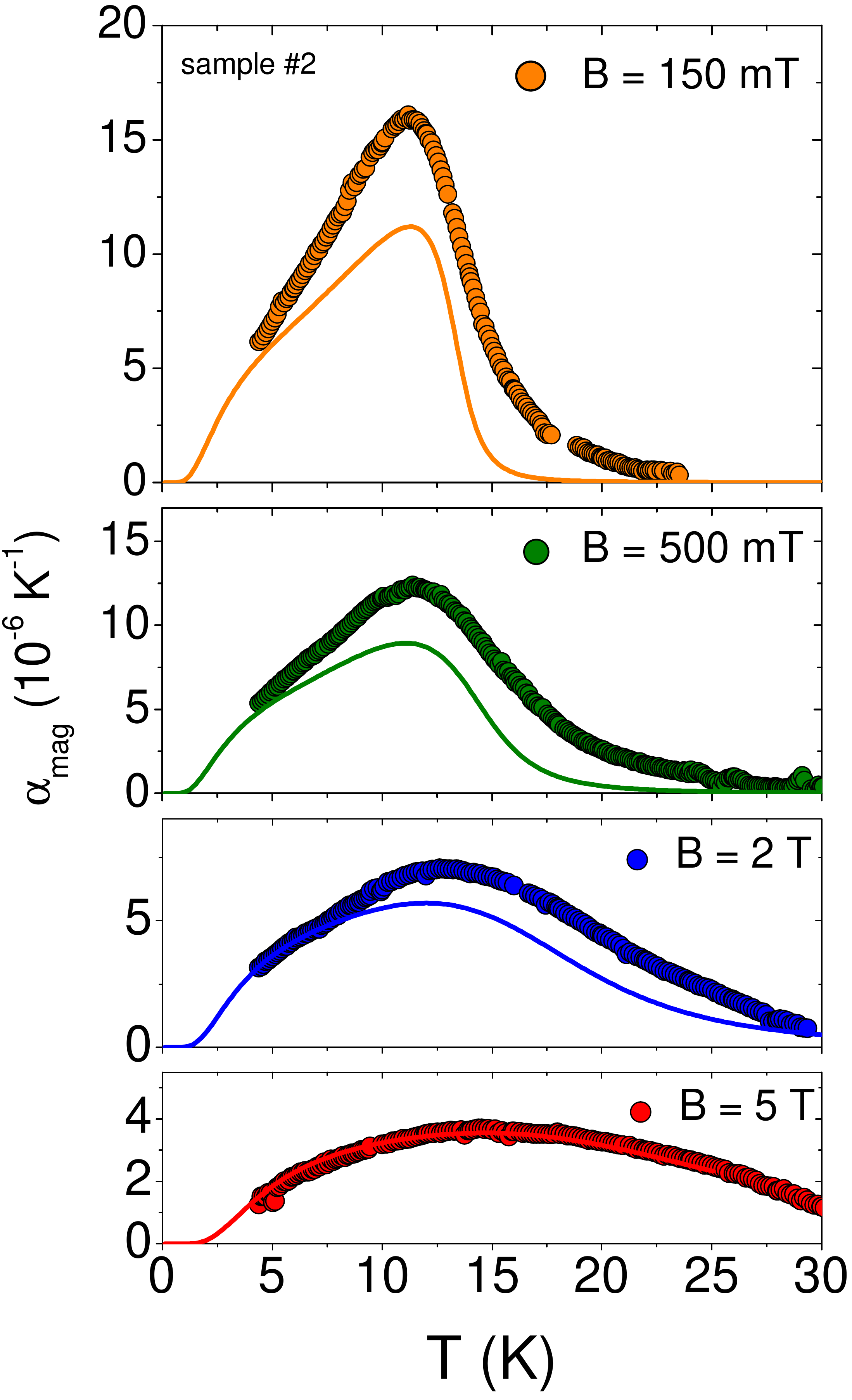}
\caption{\label{Fig2}(color online). Magnetic field dependence of the magnetic contribution to the thermal expansion coefficient $\alpha_{\rm mag}(T) = \alpha_{\rm EuB_6}(T) - \alpha_{\rm ph}(T)$, where a small phonon part is subtracted. Solid lines are mean-field calculations for a Heisenberg ferromagnet  with $S = 7/2$ and $T_C = 12.6$\,K.}
\end{figure}
\begin{figure}[htb]
\includegraphics[width=1\columnwidth]{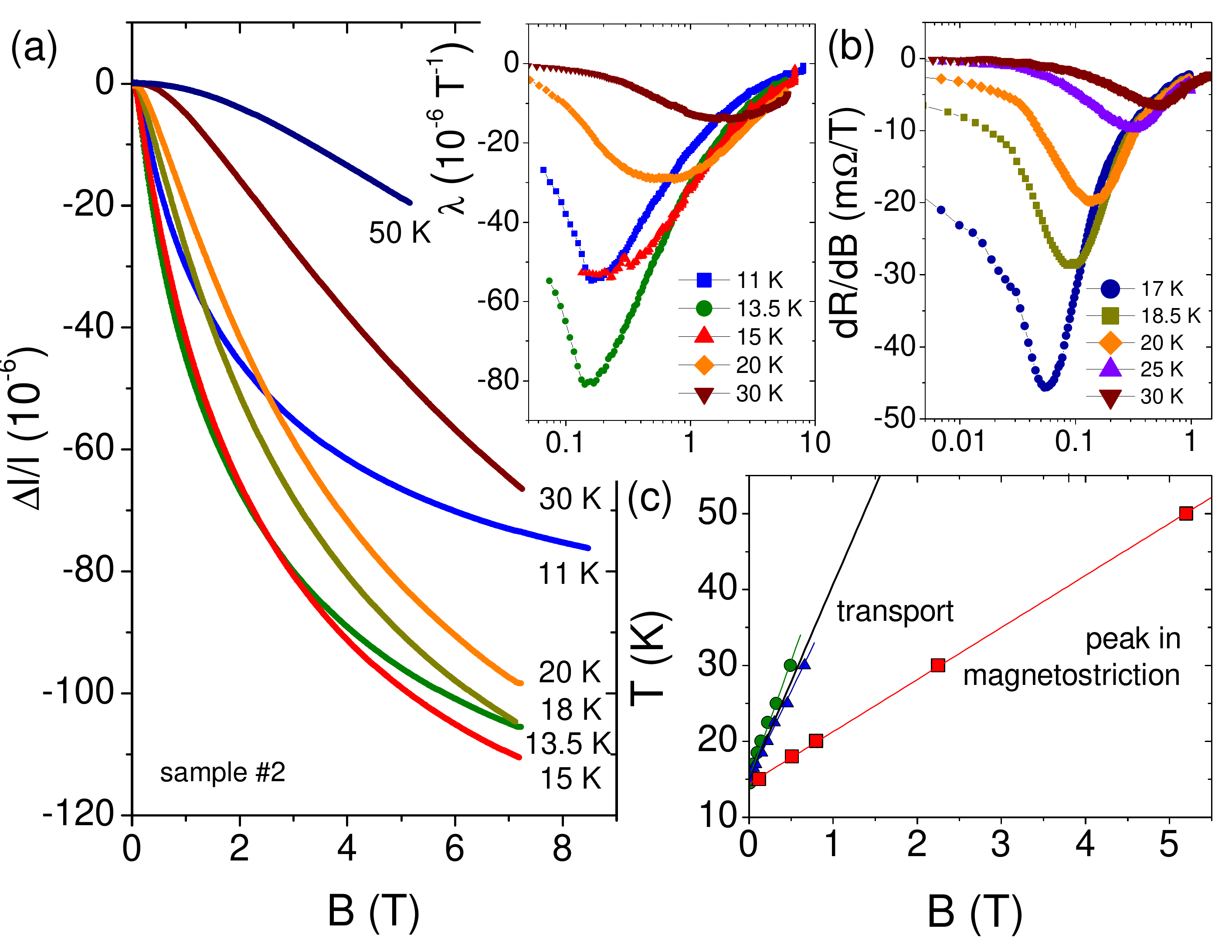}
\caption{\label{Fig3}(color online). (a) Relative length change as a function of applied magnetic field, $\Delta l/l$ {\it vs.}\ $B$, at different constant temperatures. Inset shows the corresponding magnetostriction coefficient $\lambda = \partial \ln{l}/\partial B$. (b) Magnetic field derivative of the magnetoresistance  at various temperatures.  (c) $B$-$T$ phase diagram showing the peaks in the magnetostriction coefficient $\lambda$ (red squares) with a linear fit (red line) to the data. Also shown is the carrier delocalization transition as determined from the switching field of the Hall resistivity (black line) \cite{ZhangPRL2009}, which nearly coincides with the peak in the temperature derivative of the magnetoresistance (green circles), see (b), and a peak in nonlinear transport (blue triangles) \cite{Amyan2013}, which were measured on different samples.}
\end{figure}

Further evidence for an anomalous contribution to the lattice strain can be found by studying the effect of a magnetic field on $\alpha(T)$ at low temperatures, shown in Fig.\,\ref{Fig2} for a selection of magnetic fields. Whereas the transition at $T_{c_1}$ is strongly influenced by the magnetic field \cite{comment1} 
the pronounced peak associated with $T_{c_2}$ is much less field dependent. 
With increasing field, the peak becomes suppressed in magnitude and develops into a rounded maximum with a progressively broadened high-temperature tail, whereas its position slightly shifts to higher temperatures.
For discriminating an anomalous lattice effect from an ordinary exchange striction-type contribution of local-moment magnetic ordering, we compare the data with model calculations for a Heisenberg ferromagnet with nearest neighbour exchange $J$. Within mean-field (MF) theory, a magnetic contribution ($\Delta l/l)^{\rm MF}_{\rm mag} \propto M^{2}$ is expected, with $M$ the magnetization, see \cite{Rodriguez2005,Heyer2011} and references cited therein. This results in
\begin{equation}
\alpha^{\rm MF}_{\rm mag}(T,B) = c  M(T, B) \frac{\partial M(T, B)}{\partial T}.
\end{equation}
Here $M(T, B)$ is obtained from solving the Brillouin function for $S = 7/2$ at various magnetic fields $B$ and $c$ is a constant proportional to the magnetic Gr\"uneisen parameter $\Gamma_{\rm mag}$.
As shown by the solid lines in Fig.\,\ref{Fig2}, this model, with $c$ as the only adjustable parameter, provides an excellent description of the data at 5\,T for $T \leq$ 30\,K. At this magnetic field level, MPs are widely suppressed as deduced, {\it e.g.}, from magnetoresistance measurements \cite{SuellowPRB1998}, and a homogeneous magnetic state develops. By using the same constant $c$ \cite{Heyer2011}, parameter-free model curves can then be calculated for all other fields depicted in Fig.\,\ref{Fig2}. The figure shows that these curves for fields $B < 5$\,T provide a reasonably good description of $\alpha_{\rm mag}$ only at its low-temperature end, {\it i.e.}, sufficiently deep in the FM state, but deviate considerably at intermediate temperatures and for the high-temperature tail. For small fields $B \lesssim 500$\,mT, these deviations are particularly strong around $15\,{\rm K}$, corresponding to $T_{c_1}(B = 0)$, below which MPs percolate, {\it i.e.}, where also the CMR effect is largest \cite{SuellowPRB2000,Urbano2004}.\\
Such an interrelation of an anomalous lattice strain and the CMR effect for $T_{c_2} < T \lesssim T_{c_1}$ is corroborated by magnetostriction measurements. In Fig.\,\ref{Fig3}(a) we show the $B$-induced relative length change 
at various constant temperatures in the FM and PM regimes. The data in the FM regime, {\it e.g.}, at $T$ = 11\,K, reveal a considerable $B$-induced contraction, well-accounted for by local-moment FM ordering in the mean-field model.
However, a distinctly stronger effect is observed by slightly increasing the temperature to 13.5\,K, 
{\it i.e.}, between $T_{c_1}$ and $T_{c_2}$, the polaronic percolation regime. Remarkably, $\Delta l/l$ {\it vs.}\ $B$ is largest at $T \sim 15$\,K, {\em i.e.}, close to $T_{c_1}(B = 0)$, where also the CMR effect is largest. These observations, together with the strikingly similar shapes of the magnetostriction and differential magnetoresistance d$R$/d$B$ curves in Fig.\,\ref{Fig3}(b), suggest a close interrelation between lattice strain and field-induced charge carrier delocalization. Figure\,\ref{Fig3}(c) compiles in a $B$-$T$ diagram the positions of the minima in $\lambda(B, T = {\rm const.})$ and ${\rm d}R/{\rm d}B(B, T = {\rm const.})$ and anomalies in other transport properties reported in the literature: the signatures of charge delocalization in the PM regime, as deduced from Hall effect \cite{ZhangPRL2009}, the anomalies in linear and nonlinear magnetotransport \cite{Das2012,Amyan2013}, as well as the position of the minimum in $\lambda$ (this work), all exhibit an in-$B$-linear behavior, which for $B \rightarrow 0$ extrapolates to the {\em upper} transition $T_{c_1}$, related to the charge delocalization. 
Since this temperature also coincides with the PM Curie temperature $\Theta_p$, the anomalies in the various quantities occur at a constant value of $B/(T - \Theta_p)$, which in turn is proportional to the magnetization in the PM phase (Curie-Weiss law).

The above measurements on EuB$_6$ reveal -- in addition to the ordinary exchange-driven lattice effects due to FM order -- anomalous contributions to the lattice strain both as a function of $T$ and $B$, which can be assigned to the stabilization and subsequent percolation of MPs. 
Upon cooling in zero field, this anomalous lattice strain sets in at $T^\ast$, where bound MPs become stabilized.
This may be explained by local lattice distortions surrounding these isolated objects caused by the Coulomb effects from the surrounding point charges of the lattice acting on them.
This effect, which describes the influence of fourth-order CEF splitting on the variation of the exchange constant between neighboring Eu$^{2+}$ ions with the lattice parameters, has been estimated in Ref.\ \cite{GuySSC1980}. It is found that this effect alone {\em cannot} explain the large Gr\"uneisen parameter at the low-temperature transitions $T_{c_1}$ and $T_{c_2}$, in agreement with our interpretation of a substantial contribution from charge delocalization described below. Assuming a nearest-neighbor point-charge model with a charge of $-2$ at the ligand site \cite{Turberfield1970}, however, a rough estimate using the parameters given in \cite{GuySSC1980} yields $\Gamma_{\rm eff} \sim 4 - 6$, which indeed is observed at around $T = 20$\,K.
Upon cooling, the MPs grow in size (and possibly number) 
until at $T_{c_1}$ the percolation threshold is reached and the holes become delocalized. This sudden increase of 'metallicity' goes along with the formation of an infinite magnetic cluster, involving the ordering of $15 - 20\,\%$ of the magnetic moments as the percolation progresses until $T_{c_2}$ is reached, where the magnetic clusters merge.
Below $T_{c_2}$, the process of charge delocalization levels off and spontaneous magnetization due to FM exchange prevails. This substantial lattice strain accompanying the magnetically-driven delocalization of charges is similar to what is observed across the Mott metal-insulator transition in molecular conductors \cite{deSouza2007}. In both cases, the charge delocalization, which strengthens the chemical binding, may account for the pronounced lattice contraction.\\
For finite magnetic fields, an additional contribution to the expansivity is observed in the PM temperature region, see Fig.\ \ref{Fig2}. The coincidence of the temperatures, where both the magnetostriction and the CMR effect are largest, see Fig.\ \ref{Fig3}, suggests a common origin, namely magnetically-driven charge delocalization. The fact that in the PM regime this effect is a crossover rather than a phase transition may explain why the anomalies in transport and thermodynamic properties occur at different values of a critical magnetization.

So far, in Eu-chalcogenide alloys and magnetic semiconductors, and also for the present EuB$_6$, MPs have been assumed to be unaccompanied by lattice distortions \cite{Molnar1967,Kasuya1968,Molnar1983}.
However, our results highlight a close interrelation of transport, magnetic and elastic properties. Comparing these findings for a simple material like EuB$_6$, which is devoid of additional JT lattice distortion or strong CEF effects, with the observations for the manganites and other CMR systems suggests that a strong effect of the formation and dynamics of magnetic clusters on the lattice parameters is a general feature of CMR materials.



\begin{thebibliography}{99}
\bibitem{comment01} We neglect the very small linear contribution of free charge carriers.
\end{thebibliography}
\end{document}